\documentclass{sig-alternate-2013}

\usepackage{type1cm}     
\usepackage{graphicx}     
\usepackage{wrapfig}	
\usepackage{xspace}     
\usepackage{balance}     
\usepackage{booktabs}     
\usepackage{multi row}     
\usepackage[font={bf}, tableposition=top]{caption}     
\usepackage{bold-extra}     
\usepackage{siunitx}          
\usepackage[vlined,linesnumbered,ruled,noend]{algorithm2e}     
\usepackage[bookmarks, pdftex, colorlinks=false]{hyperref}     
\usepackage[square,numbers]{natbib}     
\usepackage{microtype}    
\usepackage{units}     
\usepackage{mathtools}     
\usepackage{amssymb}     
\usepackage[show]{chato-notes}
\usepackage{relsize}
 \usepackage{graphicx}
\usepackage{caption}
\usepackage{booktabs}     
\usepackage{multirow}
\usepackage{mathtools} 
\usepackage[]{inputenc}
\usepackage{amssymb}
\usepackage{amsmath}
\usepackage{siunitx} 
\usepackage{xfrac}
\usepackage[]{algorithm2e}
\usepackage[font={small}]{subfig} 

\makeatletter
\newcommand\footnoteref[1]{\protected@xdef\@thefnmark{\ref{#1}}\@footnotemark}
\makeatother

\newcommand{\spara}[1]{\smallskip\noindent\textbf{#1}}

\newcommand{\para}[1]{\noindent\textbf{#1}}

\permission{\copyright 2017 International World Wide Web Conference Committee \\ (IW3C2), published under Creative Commons CC BY 4.0 License.}
\conferenceinfo{WWW'17 Companion,}{April 3--7, 2017, Perth, Australia.}
\copyrightetc{ACM \the\acmcopyr}
\crdata{978-1-4503-4914-7/17/04. \\
http://dx.doi.org/10.1145/3041021.3054737 \\
\includegraphics{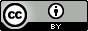}
}

\begin{document}


\title{Exposing Twitter Users to Contrarian News\thanks{Accepted as a demo at WWW 2017. Please cite the WWW version.}}

\numberofauthors{4}


\author{
\end{tabular}\begin{tabular}[t]{p{0.4\textwidth}}\centering
Kiran Garimella\\
       \affaddr{Aalto University}\\
       \affaddr{Helsinki, Finland}\\
       \affaddr{kiran.garimella@aalto.fi}
\end{tabular}\begin{tabular}[t]{p{0.4\textwidth}}\centering
Gianmarco De~Francisci~Morales\\
      \affaddr{Qatar Computing Research Institute}\\
      \affaddr{Doha, Qatar}\\
      \affaddr{gdfm@acm.org}
\and 
\end{tabular}\begin{tabular}[t]{p{0.4\textwidth}}\centering
Aristides Gionis\\
      \affaddr{Aalto University}\\
      \affaddr{Helsinki, Finland}\\
      \affaddr{aristides.gionis@aalto.fi}
\end{tabular}\begin{tabular}[t]{p{0.4\textwidth}}\centering
Michael Mathioudakis\\
      \affaddr{Aalto University \& HIIT}\\
      \affaddr{Helsinki, Finland}\\
      \affaddr{michael.mathioudakis@aalto.fi}
}

\maketitle

\begin{abstract}
Polarized topics often spark discussion and debate on social media. 
Recent studies have shown that polarized debates have a specific clustered structure in the endorsement network, which indicates that users direct their endorsements mostly to ideas they already agree with.
Understanding these polarized discussions and exposing social media users to content that broadens their views is of paramount importance.

The contribution of this demonstration is two-fold.
($i$) A tool to visualize retweet networks about controversial issues on Twitter.
By using our visualization, users can understand how polarized discussions are shaped on Twitter, and explore the positions of the various actors. 
($ii$) A solution to reduce polarization of such discussions.
We do so by exposing users to information which presents a contrarian point of view.
Users can visually inspect our recommendations and understand why and how these would play out in terms of the retweet network.

Our demo\footnote{\small\url{https://users.ics.aalto.fi/kiran/reducingControversy/homepage}} provides one of the first steps in developing automated tools that help users explore, and possibly escape, their echo chambers.
The ideas in the demo can also help content providers design tools to broaden their reach to people with different political and ideological backgrounds.

\end{abstract}

\section{Introduction}

Social media provides a platform for public discussions on a wide range of topics.
Even though users are potentially given a vast choice on what information they can consume, echo chambers, combined with social network design, often limit users to viewpoints that they agree with.
Particularly for controversial topics, discussions on social media tend to become polarized, with users supporting their stance and ignoring the one from the opposing side.

Here we present an interactive demo that ($i$) showcases the phenomenon of polarization for discussions on controversial topics, and ($ii$) provides \emph{contrarian} content recommendations, i.e., content that expresses views from the opposing side of the controversy.\footnote{\small\url{https://en.wikipedia.org/wiki/Mary,_Mary,_Quite_Contrary}}
The goal of this demo is to allow users to explore polarized discussions on social media and, at the same time, show a way to address the polarization phenomenon.

The experience focuses on controversial topics, and especially on the way information is disseminated in these discussions.
As previous studies have confirmed, controversial topics create more polarized discussions, which are characterized by specific types of sharing and information dissemination patterns~\cite{garimella2016quantifying}.
We model the discussion as an endorsement graph (e.g., retweets in Twitter).
In this graph there is an edge between two users $u$ and $v$ if $u$ retweets $v$.

These polarized discussions, in which people reinforce their existing beliefs, lead to the creation of echo chambers and filter bubbles.
Many studies have warned about the threats that these phenomena pose to an open democratic process, as they cultivate isolation and misunderstanding across sections of the society~\cite{sunstein2009republic}.

Our demonstration shows a possible way to address this problem, by exposing people to contrarian content.
Differently from previous attempts, our system is fully automated, and employs an algorithm to recommend a set of contrarian news articles that have been shared by the opposing side.
By exposing users to content which supports contrarian beliefs, we hope to encourage people to look and understand the point of view of the other side of the controversy.

\enlargethispage{-\baselineskip}

Our recommendations take into account several factors.
The main one comes from recent research in connecting users with opposing views~\cite{garimella2017reducing}.
This factor quantifies the reduction in polarization that a successful recommendation (a recommendation that is endorsed by the recipient) would generate in the discussion.

Clearly, many users might not be interested in content from the other side. 
For instance, people in the `center' might be more eager to explore content from either side, as opposed to people on the extremes.
To address this concern, our algorithm takes into account each user's history, by computing how likely they are to endorse our recommendation.

The recommendations also take other factors into account, such as the topic distribution and popularity of the content.
These factors allow users to get more diverse and engaging content recommendations.


We use several controversial topics ranging over the last two years across multiple domains for this demo, though in practice any controversial topic can be easily incorporated into our demo.

\section{Related Work}

Even though the Web was envisioned as a place where open discussions on wide range of topics could be facilitated, many users currently do not make use of such an opportunity. 
Due to phenomena such as homophily, confirmation bias, and selective exposure, people tend to restrict themselves to viewing and sharing information that conforms with their beliefs.
Research shows that this phenomenon exists online~\cite{liao2013beyond}.
This selective exposure has led to increased fragmentation and polarization online. 
A wide body of recent studies have studied~\cite{adamic2005political,conover2011political,mejova2014controversy} and 
quantified~\cite{akoglu2014quantifying,garimella2016exploring,garimella2016quantifying,guerra2013measure,morales2015measuring} 
this issue.

Research also shows that such a division in sections of the society has important consequences to our democracy~\cite{sunstein2009republic}.
There have been attempts to try to nudge users to explore and understand opposing view points.
Here we review the most relevant ones.

Wall Street Journal's \emph{Blue feed-Red feed}~\footnote{\small\url{http://graphics.wsj.com/blue-feed-red-feed/}} raises awareness about the extent to which viewpoints on a matter can differ, by showing side-by-side articles expressing very liberal and very conservative viewpoints;
{\it Politecho}\footnote{\small\url{http://politecho.org/}} displays how polarizing the content on a user's news feed is when compared to their friends';
{\it Escape your bubble}\footnote{\small\url{http://www.escapeyourbubble.com/}} 
is a browser extension to add hand-curated content from the opposite side on Facebook;
automated bots have been created to respond to posts containing harassment or fake news,\footnote{\small\url{http://wpo.st/4kVR2} \url{https://goo.gl/Xl6x9t}} with an attempt to de-polarize the discussion and educate users.
Moreover, new social media platforms designed to encourage discussions and debates have been proposed, such as 
($i$) the Filterburst project,\footnote{\small\url{http://www.filterburst.com/}}
($ii$) Rbutr,\footnote{\small\url{http://rbutr.com/}} where users can post rebuttals of other urls, and 
($iii$) a Wikipedia for debates.\footnote{\small\url{http://www.debatepedia.org/en/index.php/Welcome_to_Debatepedia\%21}}

Our demo differs from existing ones in many ways. 
First, we provide a unique, interactive visualization of an endorsement networks for controversial topics.
Second, we showcase a system to recommend contrarian content to users. Our approach is completely algorithmic, 
unlike most systems listed above, which involve manual curation.

Research has also been done in trying to connect users with opposing views.
\citet{munson2013encouraging} created a browser widget that measures 
the bias of users based on the news articles they read.
Their study shows that users are willing to slightly change views 
once they are shown their biases.
\citet{graells2013data} show that mere display of contrarian content has negative emotional effect.
To overcome this effect, they propose a visual interface for making
recommendations from a diverse pool of users, 
where diversity is with respect to user stances on a topic.
\citet{graells2014people} propose to find topics that may be of interest to both sides 
by constructing a topic graph. They define intermediary topics to be those topics 
that have high betweenness centrality and topic diversity.
\citet{park2009newscube} propose methods for presenting multiple aspects of news to reduce bias.
\citet{garimella2017reducing} study the problem of reducing the overall polarization of a controversial topic 
in a network. They try to find the best edges to recommend in an endorsement graph so that the polarization score of the entire network is reduced.
In this demo, our focus is on reducing the polarization of an individual user (local objective), instead of the entire network (global objective).

\section{Preliminaries}
\label{sec:preliminaries}
A topic of discussion is identified as the set of tweets that satisfy a text query -- e.g., all tweets that contain a specific hashtag.
We represent a topic with an \emph{endorsement graph} $G(V, E)$, 
where vertices $V$ represent users
and edges $E$ represent \emph{endorsements}.

It has been shown that
an endorsement graph captures well the extent to which 
a topic is controversial~\citep{garimella2016quantifying}.
In particular, 
the endorsement graph of a controversial topic 
has a \emph{multimodal clustered structure}, where each cluster of vertices
represents one viewpoint on the topic.
As we focus on two-sided controversies, we identify the two sides of a controversial topic by employing a \emph{graph-partitioning} algorithm, 
which partitions the graph into \emph{two} subgraphs (represented by $X$ and $Y$).
In this work, we specifically focus on recommending content in the form of news items, 
such as articles, blog posts, and opinion pieces.
The item pool for the recommendation comprises all the links shared by the active users 
during the observation window.

\spara{User polarization score.}
We use a recently-proposed metho\-do\-logy 
to define the polarization score for each user in the graph~\cite{garimella2016tscpaper}.
The score is based on the expected hitting time of a random walk that starts 
from the user under consideration and ends on a high-degree vertex on either side.
Typically in a retweet graph, 
high degree vertices on each side are indicators of authoritative content generators.
We denote the set of the $k$ highest degree vertices on each side by $X^+$ and $Y^+$.
Intuitively, a vertex is assigned a score of higher absolute value (closer to $+1$ or $-1$), 
if, compared to other vertices in the graph, 
it takes a very different time to reach a high-degree vertex on either side ($X^+$ or $Y^+$) (in terms of information flow). 
Specifically, for each vertex $u\in V$ in the graph, 
we consider a random walk that starts at $u$, and 
estimate the expected number of steps, $l_u^{_X}$ 
before the random walk reaches any high-degree vertex in $X^+$.
Considering the distribution of values of $l_u^{_X}$ across all vertices $u \in V$, 
we define $\rho^{_X}(u)$ as the fraction of vertices $v\in V$ with $l_v^{_X} < l_u^{_X}$.
We define $\rho^{_Y}(u)$ similarly.
Obviously, we have $\rho^{_X}(u), \rho^{_Y}(u) \in [0,1)$.
The polarization score of a user is then defined as
\begin{equation}
	\rho(u) = \rho^{_X}(u) - \rho^{_Y}(u) \quad \in (-1, 1) .
\end{equation}
Following this definition, a vertex that is close to high-degree vertices $X^+$, 
compared to most other vertices, will have $\rho^{_X}(u) \approx 1$;
on the other hand, if the same vertex is far from high-degree vertices $Y^+$, 
it will have $\rho^{_Y}(u) \approx 0$;  
leading to a polarization score $\rho(u) \approx 1 - 0 = 1$.
The opposite is true for vertices that are far from $X^+$ but close to $Y^+$; 
leading to a polarization score $\rho(u) \approx -1$.

\spara{Item polarization score.}
Once we have obtained polarization scores for users in the graph, it is straightforward to derive a similar score for content items shared by these users.
Specifically, we define the polarization score of an item $i$ as the average of the polarization scores of the set of users who have shared $i$, denoted by $U_i$:
\begin{equation}
	\rho(i) = \frac{\displaystyle\sum_{u \in U_i} \rho(u) }{|U_i|} \quad \in (-1, 1) .
\end{equation}

\spara{Acceptance probability.}
Not all recommendations are agreeable, 
especially if they do not conform to the user's beliefs.
To reduce these effects, we define an acceptance probability, 
which quantifies the degree to which a user is likely to endorse the recommended content.
We use the item and user polarization scores defined above to estimate the likelihood that a target user $u$ endorses (i.e., retweets) the recommended item $i$.
We build an acceptance model by adapting a similar one based 
on the feature of user polarization~\citep{garimella2017reducing}.
High absolute values of user polarization
(close to $-1$ or $1$) indicate that the user belongs clearly to one side of the controversy, 
while middle-range values (close to $0$) indicate that the user is in-between the two sides.
It had been shown that users
accept content from different sides with different probabilities, and 
 these probabilities can be inferred from the graph structure~\cite{garimella2017reducing}.
For example, a user with polarization close to $-1$ is more likely to endorse a user with a negative polarization than a user with polarization $+1$. 
This intuition directly translates to endorsing items, and therefore can be used for our recommendation problem.

Based on this intuition, we define the acceptance probability $p(u,i)$ 
of a user $u$ endorsing item $i$ as
\begin{equation}
p(u,i) = N_{e}(\rho(u),\rho(i))/N_{x}(\rho(u),\rho(i)),
\end{equation}
where $N_{e}(\rho(u),\rho(i))$ and $N_{x}(\rho(u),\rho(i))$ are the number of times a user with polarity $\rho(u)$ has endorsed and was exposed to (respectively) 
content of polarity $\rho(i)$. 
In practice, the polarity scores are bucketed to smooth the probabilities.

\section{Recommendation factors}
\label{sec:recommendation_factors}

This section describes the factors used to generate content recommendations for the users.
Though our main focus is to connect users with content that expresses a contrarian point of view, we also want to maximize the chances of such a recommendation being endorsed by the user.
%
Therefore, we take into account several factors: reduction in polarization score of the target user;
exclusivity of the candidate items (polarity of the items);
acceptance probability of recommendation based on polarization scores;
topic diversity;
popularity/quality of the candidate item.
Next, we describe these factors in more detail.

\spara{Reduction of user-polarization score.}
The maximum reduction of user-polarization score is achieved by putting the user in contact 
with an authoritative source from the opposing side.
Leveraging this idea, we build a list of items $L_1$, 
by considering the items shared by high degree nodes on the opposite side of the target user, 
and ranking them by the potential decrease in polarization score for a user $u$. 

\spara{Exclusivity on either side.}
We consider items that are almost exclusively shared by one of the sides.
Specifically, we denote by $n_i^X$ and $n_i^Y$ the number of users 
who shared each item $i$ on side $X$ and $Y$, respectively.
For each side, we generate a list $L_2$ ranked 
by the ratio of shares $\sfrac{n_i^X}{n_i^Y}$ (for side $X$) and $\sfrac{n_i^Y}{n_i^X}$ (for side $Y$).

\spara{Acceptance probability.}
For a given user $u$, all items sorted in decreasing order of acceptance probability $p(u,i)$ make up list $L_3$.

\spara{Topic diversity.}
We want to ensure that the recommendations are topically diverse.
Therefore, for each user, we compute a vector $t_u$ 
that contains the topics extracted from the tweets written and the items shared by the user.
Similarly, we extract a vector of topics $t_i$ for each item. 
Topics are defined as {\em named entity}, 
and we extract them via the tool TagMe.\footnote{\url{https://services.d4science.org/web/tagme}}
Given a user vector $t_u$, we compute the cosine similarity with all item vectors $t_i$, 
and rank items in increasing order of cosine similarity (list $L_4$).

\spara{Popularity on either side.}
Finally, we take into account the popularity of the recommended items, 
so that users receive content that is popular and, likely, of good quality.
For each item, we compute a popularity score as the maximum number of retweets 
obtained by a tweet that contains this item.
List $L_5$ ranks items by decreasing popularity score.

\spara{Rank Aggregation.}
Given the 5 ranked lists described above,
we use a weighted rank-aggregation scheme to generate the final recommendations.
The intuition behind using rank aggregation is that items that are highly ranked in many lists, 
are also highly ranked in the output list.
In particular, we use a weighted rank-aggregation technique proposed by \citet{pihur2009rankaggreg}:
the goal is to minimize an objective function
\begin{equation}
	\phi(\delta) = \sum_{i=1}^{5} w_i d(\delta,L_i) ,
\end{equation}
where $\delta$ is the optimal ranked output list, $d$ is any distance function 
(we use the Spearman footrule distance), and 
$w_i$ are the importance weights of each list.
We can set the weights to generate highly contrarian recommendations 
(by giving large weights to $L_1$ and $L_2$) or 
recommendations that are likely to be accepted (by giving large weight to $L_3$).

\section{Architecture}

The demo consists of three major parts. (i) Data collection, (ii) Data processing, creation of graphs, and recommendations (detailed in Secions~\ref{sec:preliminaries} and \ref{sec:recommendation_factors}) and (iii) Visualization. We give a brief overview of each of these below.

\spara{Data Collection.}
We collected data from Twitter for eight controversial topics covering a wide range of domains, including the US election results (USElections), and protests against government actions (\#baltimoreriots, \#beefban, and \#nemtsov). Each of these topics contain a few thousands to tens of thousands of users.
Though we limit ourselves to these 8 topics in the demo, in practice, a similar methodology can be applied for any controversial topic.

\spara{Data Processing.}
After identifying a controversial topic on Twitter, we construct a \emph{retweet graph}, identify the two sides of the controversy and obtain polarity scores for all users in this topic (detailed in Section~\ref{sec:preliminaries}). 
Next, for each user, we generate recommendations that surface content with a contrarian point of view. The recommendations for a user are based on their Twitter activity, and take into account five different factors. The details for extracting the recommendations are provided in Section~\ref{sec:recommendation_factors}.

\spara{Visualization.}
To visualize the results in the demo we use a javascript library called Sigma.js.\footnote{\small\url{https://sigmajs.org}}
The retweet network visualizations are created first in Gephi,\footnote{\small\url{https://gephi.org}} using a force directed layout and then exported to Sigma.js via a plugin.\footnote{\small\url{https://marketplace.gephi.org/plugin/sigmajs-exporter}}

\section{Description of the Demo}

To use the demo,\footnote{\small\url{https://users.ics.aalto.fi/kiran/reducingControversy/homepage/}} the viewer first selects a topic from a set of polarized topics on the homepage.
Upon selecting a topic, the retweet network corresponding to the topic is shown to the viewer, with the two opposing sides of the controversy highlighted with different colors.
On the left there is an info box, where the viewer can find summary information about the topic such as what the discussion is about, why it is polarized, and what the two sides support.
The viewer can optionally get more detailed information, including most retweeted tweets on each side, most shared news articles, and other aggregate statistics on the topic, by clicking on the `More information about this visualization' link.
%
Figure~\ref{fig:screenshot1} shows the main Web interface for the discussion around the protests for the assassination of Boris Nemtsov in Russia.\footnote{\small\url{https://en.wikipedia.org/wiki/Boris_Nemtsov}}

The retweet network is at the center of the visualization. 
The retweet network for controversial topics exhibits a peculiar clustered structure, with two main clusters.
By hovering over each node, the viewer can see which other nodes they are connected to.
In most cases, nodes are connected to a single side (nodes of the same color), and connections across sides (nodes of different color) are rare.
This pattern is an indication that users do not retweet across different sides of the discussion, but only support their own point of view~\cite{garimella2016quantifying}.
The viewer can zoom in and out to see specific connections between individual users and groups.
Hovering on a node shows their Twitter username, along with their assigned polarity score (higher absolute value means that the user is more polarized). Clicking on a node in the graph highlights the subgraph connected to this node and also brings up an information pane on the right, as shown in Figure~\ref{fig:screenshot2}.

The information pane consists of ($i$) a link to the users profile on Twitter, ($ii$) a sample of three retweets by the user (if the user has retweeted anyone), and ($iii$) three recommendations that aim to expose that specific user to a contrarian viewpoint, along with a set of three random articles. Providing two lists allows the viewer to compare our recommendations to a random baseline recommendation.
These samples can be refreshed by clicking on the node again.

For each recommendation, there is a link to show/hide a popup which contains information on \emph{why} that link has been recommended.
The popup displays the normalized weights given to the five factors that went into the ranking. 
Hovering over each recommendation highlights the nodes that have shared this article.
This visualization is useful to get an idea on what part of the network shared this article, and hence understand how the recommendation could modify the structure of the endorsement graph of the discussion.
Figure~\ref{fig:screenshot3} shows a depiction of these features.


\begin{figure}
\centering
\includegraphics[width=\columnwidth]{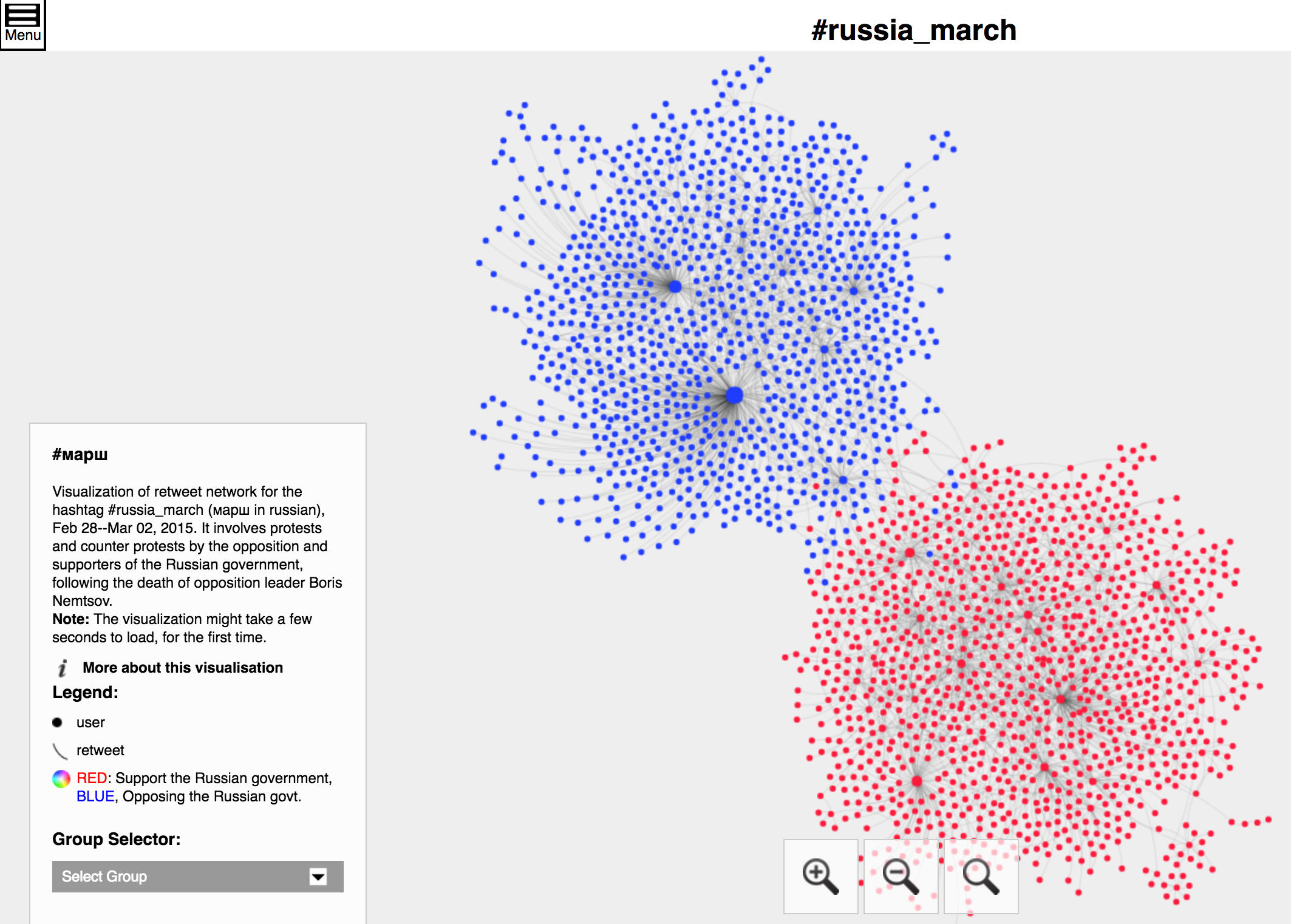}
\caption{Screenshot of the web interface for the topic \#russia\_march.}
\label{fig:screenshot1}
\end{figure}

\begin{figure}
\centering
\includegraphics[width=\columnwidth]{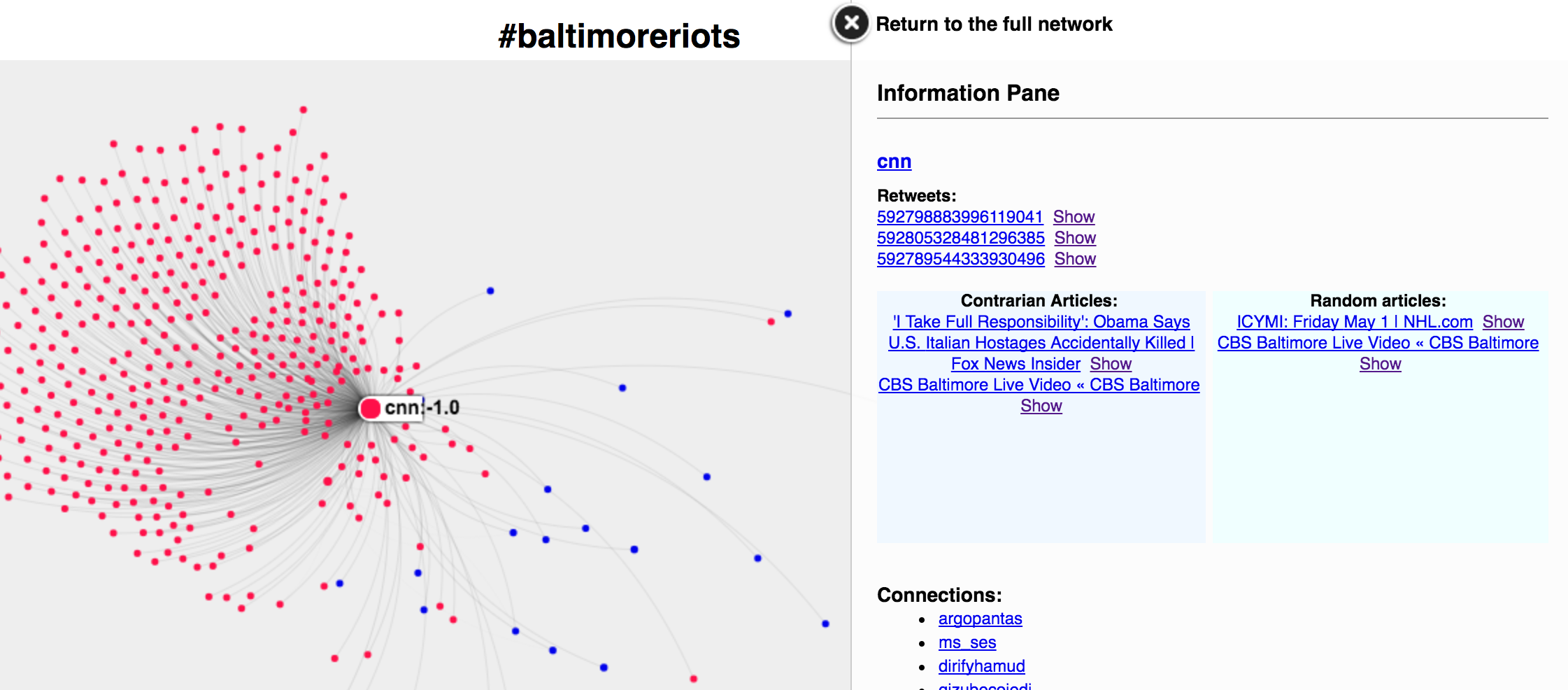}
\caption{Screenshot of the recommendations upon selecting a node (CNN) for \#baltimoreriots.}
\label{fig:screenshot2}
\end{figure}

\begin{figure}
\centering
\includegraphics[width=\columnwidth]{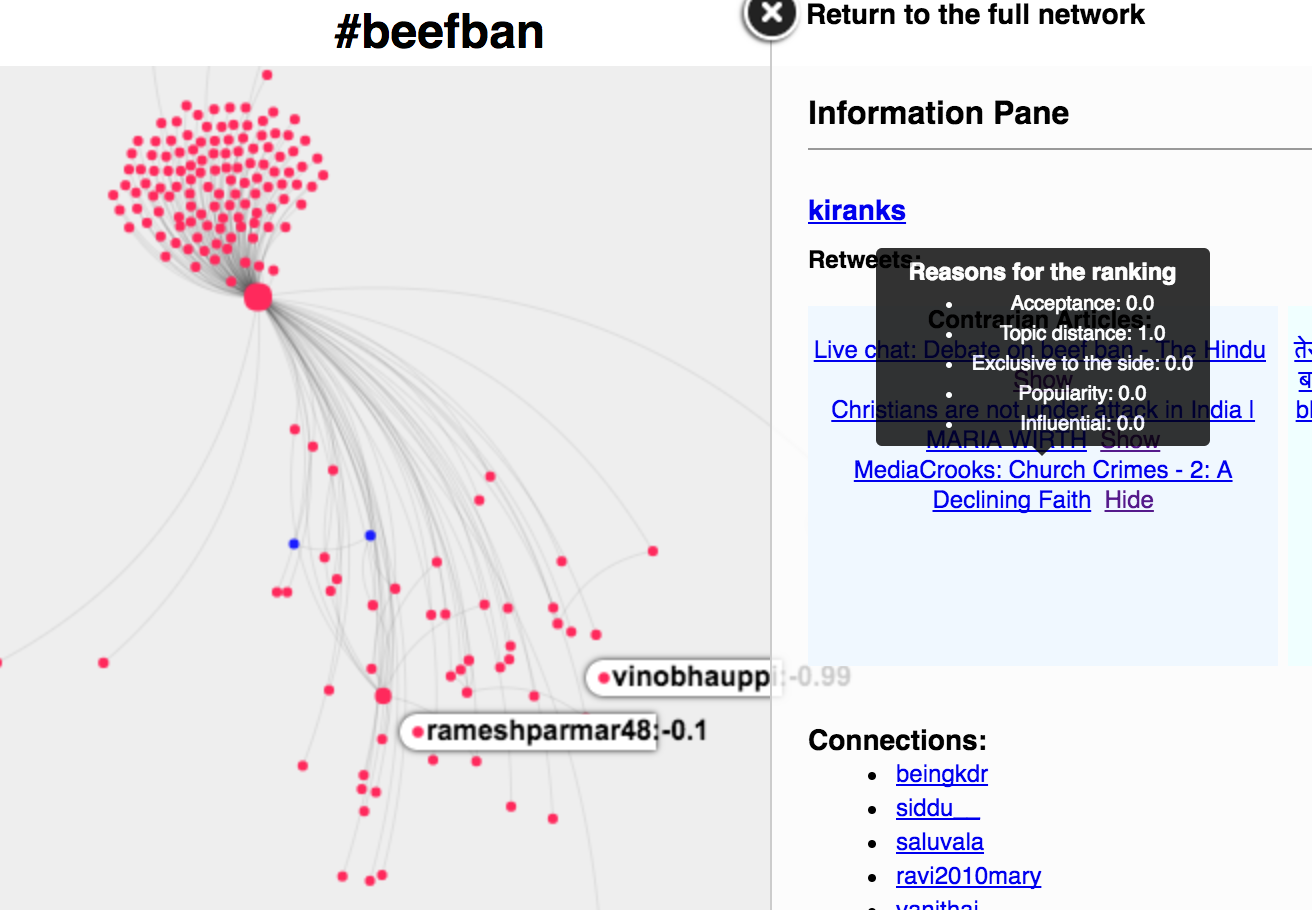}
\caption{Screenshot showing the weights for a recommendation upon selecting a node, and the users who shared the article on the graph.}
\label{fig:screenshot3}
\end{figure}

\spara{Acknowledgements.}
This work has been supported by the Academy of Finland project ``Nestor'' (286211) and the EC H2020 RIA project ``SoBigData'' (654024).

\section{Authors}

\begin{wrapfigure}{l}{0.1\textwidth}
    \includegraphics[width=0.12\textwidth]{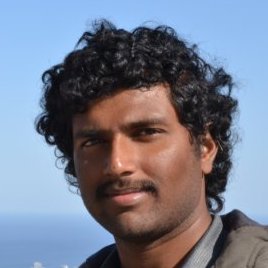}
\end{wrapfigure}
\para{Kiran Garimella} is a PhD student at Aalto University. 
Previously he worked as a Research Engineer at Yahoo Research, Qatar Computing Research Institute and as an intern at LinkedIn and Amazon.
His PhD thesis focuses on identifying, quantifying and combating filter bubbles on social media.
\\ \bigskip

\begin{wrapfigure}{l}{0.1\textwidth}
    \includegraphics[width=0.12\textwidth]{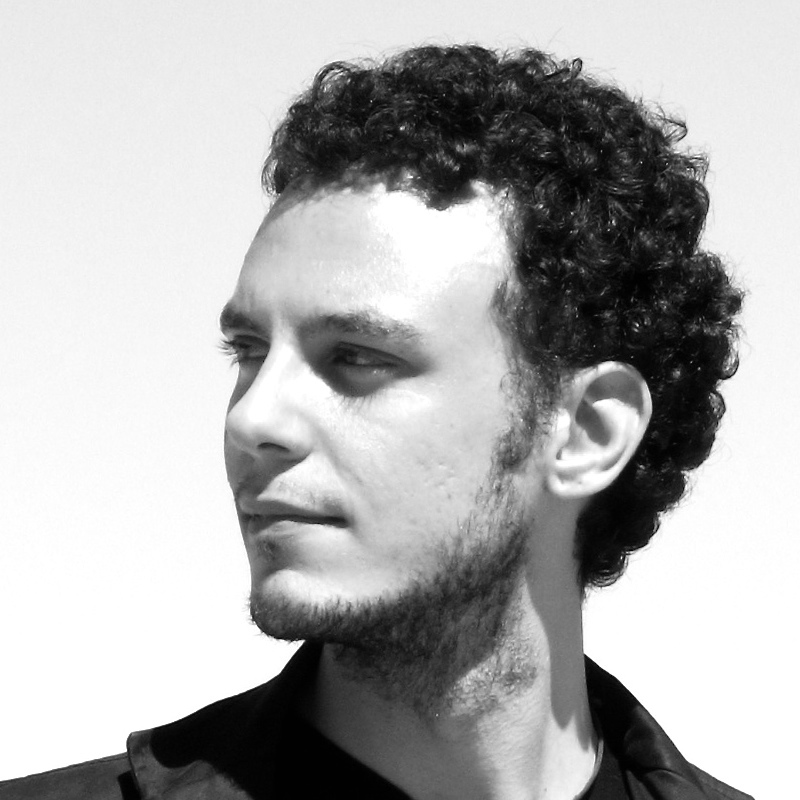}
\end{wrapfigure}
\para{Gianmarco De Francisci Morales} is a Scientist at QCRI. Previously he worked as a Visiting Scientist at Aalto University, as a Research Scientist at Yahoo Labs Barcelona, and as a Research Associate at ISTI-CNR in Pisa.
He received his Ph.D. in Computer Science and Engineering from the IMT Institute for Advanced Studies of Lucca in 2012.
His research focuses on scalable data mining, with an emphasis on Web mining and data-intensive scalable computing systems.
He is one of the lead developers of Apache SAMOA, an open-source platform for
mining big data streams.
He co-organizes the workshop series on Social News on the Web (SNOW), co-located with the WWW conference.
\bigskip

\begin{wrapfigure}{l}{0.1\textwidth}
    \includegraphics[width=0.12\textwidth]{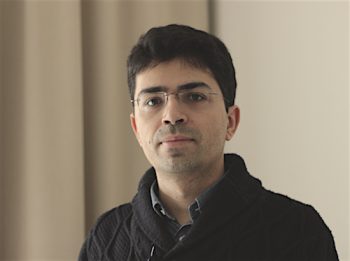}
\end{wrapfigure}
\para{Michael Mathioudakis} is a Postdoctoral Researcher at Aalto University.
He received his PhD from the University of Toronto.
His research focuses on the analysis of user generated content on social media, with a recent emphasis on urban computing and online polarization.
At Aalto University, he organized and taught new courses on `Modern Database Systems' and `Social Web Mining'.
He also serves as advisor to Master's students and Aalto's representative at the SoBigData EU project.
Outside academia, he worked as a data scientist at Helvia and Sometrik, two data analytics companies.
\bigskip

\begin{wrapfigure}{l}{0.1\textwidth}
    \includegraphics[width=0.12\textwidth]{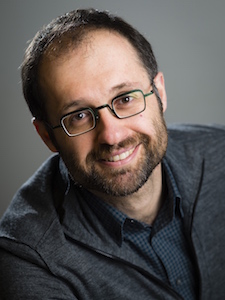}
\end{wrapfigure}
\para{Aristides Gionis} is an associate professor at Aalto University.
His research focuses on data mining and algorithmic data analysis. 
He is particularly interested in algorithms for graphs, social-network analysis, and algorithms for web-scale data.
Since 2013 he has been leading the Data Mining Group, in the Department of Computer Science of Aalto University.
Before coming to Aalto he was a senior research scientist in Yahoo! Research, and previously an Academy of Finland postdoctoral scientist in the University of Helsinki. He obtained his Ph.D. from Stanford University in 2003. 

%


\bibliographystyle{abbrvnat}
\bibliography{biblio}
\balancecolumns 

\end{document}